\documentclass[%
 reprint,
 amsmath,amssymb,
 aps,
 pra,
longbibliography,
]{revtex4-1}

\pdfoutput=1
\usepackage{graphicx}
\usepackage{hyperref}
\usepackage{braket}
\usepackage{textcomp}
\usepackage{gensymb}
\usepackage{bm}
\usepackage{relsize}
\usepackage{siunitx,xfrac}
\DeclareSIUnit{\sqrthz}{\sqrt{\hertz}}
\DeclareSIUnit{\dBm}{dBm}
\DeclareSIUnit{\dBc}{dBc}
\DeclareSIUnit{\dBi}{dBi}
\DeclareSIUnit{\Vcm}{(V/cm)}
\DeclareSIUnit{\cmV}{(cm/V)}
\DeclareSIUnit{\uVcm}{(\micro V/cm)}
\DeclareSIUnit{\uVm}{(\micro V/m)}
\DeclareSIUnit{\Vm}{(V/m)}
\DeclareSIUnit{\mVm}{(mV/m)}

\newcommand{\Erf}{{\mathcal{E}}}
\newcommand{\SNR}{{\mathrm{SNR}}}
\newcommand{\orf}{{\omega_\mathrm{RF}}}

\usepackage{xparse}
\NewDocumentCommand \state { m m m o } {%
    \IfNoValueTF {#4} {%
        \ket{#1\mathrm{#2}_{#3}}
    }{%
        \ket{#1\mathrm{#2}_{#3,m_J{=}#4}}
    }%
}

\newcolumntype{L}[1]{>{\raggedright\let\newline\\\arraybackslash\hspace{0pt}}m{#1}}
\newcolumntype{C}[1]{>{\centering\let\newline\\\arraybackslash\hspace{0pt}}m{#1}}
\newcolumntype{R}[1]{>{\raggedleft\let\newline\\\arraybackslash\hspace{0pt}}m{#1}}

\begin{document}

\preprint{APS/123-QED}
\title{Assessment of Rydberg Atoms for Wideband Electric Field Sensing}


\author{David H. Meyer}
\email[Corresponding author: ]{david.h.meyer3.civ@mail.mil}
\author{Zachary A. Castillo}
\altaffiliation[Also at: ]{Department of Physics, University of Maryland, College Park MD 20742, USA}
\author{Kevin C. Cox}
\author{Paul D. Kunz}
\affiliation{U.S. Army Research Laboratory, 2800 Powder Mill Rd, Adelphi MD 20783, USA}
\date{\today}

\begin{abstract}
Rydberg atoms have attracted significant interest recently as electric field sensors. In order to assess potential applications, detailed understanding of relevant figures of merit is necessary, particularly in relation to other, more mature, sensor technologies. Here we present a quantitative analysis of the Rydberg sensor's sensitivity to oscillating electric fields with frequencies between \SI{1}{\kilo\hertz} and \SI{1}{\tera\hertz}. Sensitivity is calculated using a combination of analytical and semi-classical Floquet models. Using these models, optimal sensitivity at arbitrary field frequency is determined. We validate the numeric Floquet model via experimental Rydberg sensor measurements over a range of \SIrange{1}{20}{\giga\hertz}. Using analytical models, we compare with two prominent electric field sensor technologies: electro-optic crystals and dipole antenna-coupled passive electronics.
\end{abstract}

\maketitle


\section{Introduction}
\label{sec:Intro}

Vapors of alkali Rydberg atoms, \emph{i.e.} where each atom's valence electron is highly excited, have recently gained attention as a promising candidate for electric field sensors thanks to some distinct characteristics.  
1) They are identical quantum particles with known response directly tied to fundamental constants.
2) They exhibit a large polarizability and sensitivity over an ultra-wide frequency range. 
3) They are small and broadly available.  
And 4) they are compatible with optical/laser technology. 
Explicit demonstrations include electric field sensitivity down to less than \SI{1}{\uVcm\per\sqrthz} \cite{kumar_rydberg-atom_2017} with record absolute accuracy \cite{holloway_broadband_2014}, detection of fields from \SI{10}{\kilo\hertz} \cite{mohapatra_giant_2008} up to \SI{1}{\tera\hertz} \cite{wade_real-time_2017}, sub-wavelength imaging \cite{fan_subwavelength_2014}, communication bandwidths of over \SI{1}{\mega\hertz} \cite{meyer_digital_2018,deb_radio-over-fiber_2018}, and effective operation in the extreme electrically small regime \cite{cox_quantum-limited_2018}.  These demonstrations provide validation for Rydberg-based sensors as a useful technology platform.

Unsurprisingly the technology space related to electric field sensing is large and varied, given the wide spectrum of frequencies and dynamic ranges that are of interest. 
More commercially mature technologies, including plasmonic sensors \cite{homola_surface_2008,schasfoort_handbook_2017}, electro-optic crystals \cite{duvillaret_electro-optic_2002}, and traditional electronic circuits coupled to antennas have found value in a broad array of marketable applications.
Other notable technologies, which can be considered quantum sensors like the Rydberg sensor, include superconducting transition edge bolometers \cite{irwin_transition-edge_2005} that have enabled cutting edge scientific results such as characterizing cosmic microwave background radiation, trapped ions \cite{brownnutt_ion-trap_2015}, and NV diamond color centers \cite{dolde_electric-field_2011,michl_robust_2019}.  
Identifying applications where the Rydberg sensor can provide a significant advantage over these technologies is an open question.

The benefits of sub-wavelength, resonant, non-destructive, precise measurements afforded by Rydberg vapors have merited applications in calibration and metrology of radio-frequency (RF) fields where current standards rely on manufactured off-resonant dipole antennas coupled to a diode rectifier \cite{holloway_broadband_2014}. 
The possibility of RF communications has been investigated recently as another potential application for Rydberg sensors \cite{deb_radio-over-fiber_2018,simons_rydberg_2019,anderson_atomic_2018,jiao_atom-based_2018,song_rydberg-atom-based_2019,meyer_digital_2018,cox_quantum-limited_2018}. However, no work so far has presented a quantitative analysis of the Rydberg sensor's sensitivity over its wide spectral range, particularly in comparison with existing electric field sensing technology. 

In this work we perform such an analysis by calculating the Rydberg sensor's field sensitivity across its operational frequency spectrum and compare with sensors of similar size (\SI{\sim1}{\centi\meter}) based on electro-optic crystals and traditional passive electronic elements. 
We begin in Section \ref{sec:Background} with a discussion of the fundamental differences in operation of these systems, followed in Section \ref{sec:QuasiDC} by analytic derivations for the sensitivity in the electrically-small, low frequency regime. 
The focus is fundamental sensitivity limits of representative model systems while highlighting various distinctive characteristics. 
In Section \ref{sec:Spectrum} we present a more generalized, numerical treatment for the Rydberg sensor in order to calculate the sensitivity for fields of arbitrary frequency up to \SI{1}{\tera\hertz}. We experimentally confirm our model's calculations for frequencies between \SIrange{1}{20}{\giga\hertz}.

\section{Background}\label{sec:Background}

For any electric field sensor, the measurement process can be divided into three stages: \textbf{1)} state preparation, including mode shaping of sensor to the incident field and/or sensor initialization, \textbf{2)} field--sensor interaction, often parameterized using macroscopic susceptibility ($\chi$) or microscopic polarizability ($\alpha$), and \textbf{3)} sensor readout. Each step impacts the various figures of merit and overall performance of a given sensor. Each step also has a fundamental limitation that depends on the type of sensor. 

When comparing disparate technologies terminology can present challenges. In particular, the notions of bandwidth and sensitivity are often used inconsistently across different communities. 

To be explicit about our terminology regarding bandwidth, ``carrier spectral range'' signifies the system's range of operational carrier frequencies, while ``instantaneous bandwidth'' signifies the maximum rate of change of the carrier to which the system is sensitive. Rydberg atoms and electro-optic crystals have a large carrier spectral range, as discussed below, while dipole-coupled passive electronics sensors are typically more restricted due to challenges of impedance matching the dipole to the readout load. In contrast, the instantaneous bandwidth for passive electronic sensors is often equal to the carrier spectral range -- resulting in little distinction between the two for this technology. The instantaneous bandwidth for electro-optic and Rydberg sensors is typically limited by the readout process. For electro-optics this corresponds to the bandwidth of the photodetector. For Rydberg sensors that rely on the electromagnetically-induced-transparency (EIT) probing method, the probe photon scattering rate of the intermediate atomic resonance (of order \SI{10}{\mega\hertz}) is the limiting bandwidth \cite{meyer_digital_2018,chen_transient_2004}.

\begin{figure*}[tb]
    \centering
    \includegraphics[width=\textwidth]{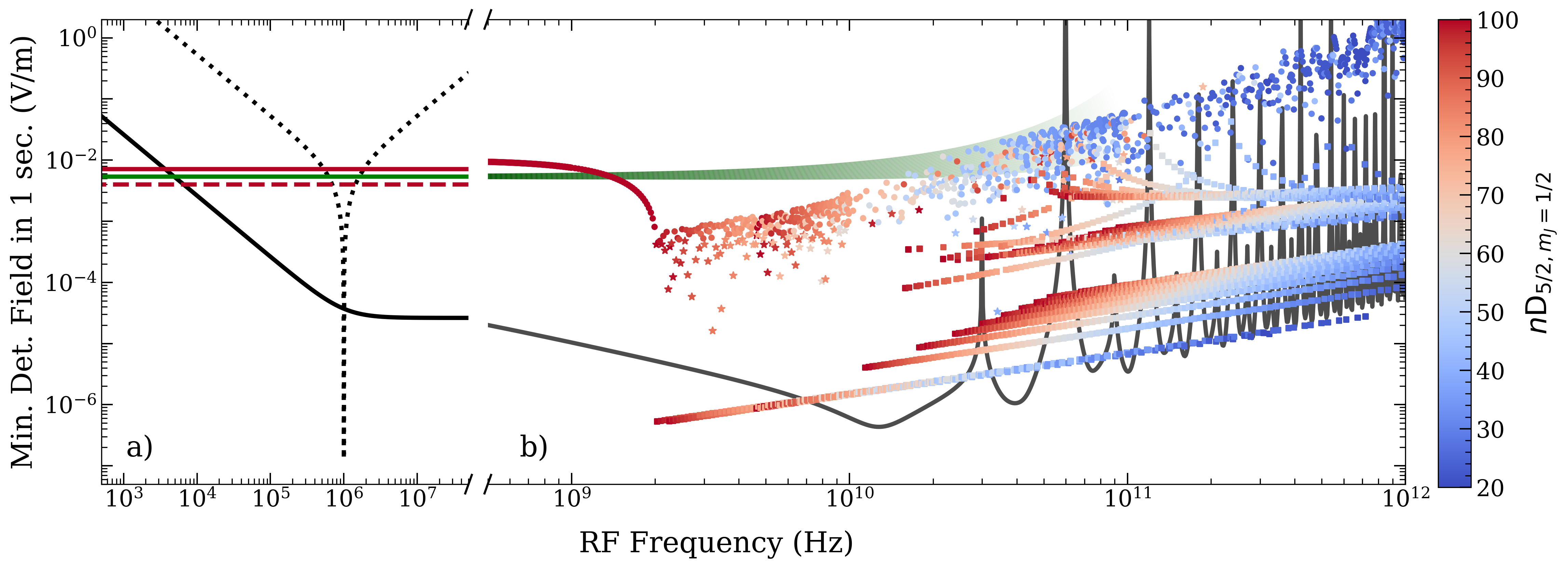}
    \caption{Minimum detectable field in a 1 second measurement versus RF frequency for \SI{1}{\centi\meter} systems.
    a) Quasi-DC regime: The solid(dashed) red lines show the minimum field using a $\state{100}{D}{5/2}[1/2]$ target state with $10^3$($10^4$) Rb  Rydberg atoms. The solid(dashed) black lines show the minimum field for a $\ell=\SI{1}{\centi\meter}$ passive dipole electronic sensor, optimized for operation at \SI{1}{\mega\hertz} with a resistive \SI{2.1}{\mega\ohm} (tuned inductor with \SI{50}{\ohm}) load. The green line shows the minimum field for a \SI{1}{\centi\meter} ZnTe electro-optic sensor with \SI{150}{\micro\watt} of optical probe power.
    b) AC-regime: Each data point represents the minimum detectable field for a Rb-based Rydberg sensor, allowing for optimal choice of $n$ denoted by color. Square, circle, and star points represent the scaling of the SNR with $\Erf$, $\beta=1,2$ or in between, respectively. The gray line shows the minimum field for a $\ell=\SI{1}{\centi\meter}$ center-fed dipole antenna terminated with a \SI{50}{\ohm} load.}
    \label{fig:PeakSensitivity}
\end{figure*}

We determine a sensor's sensitivity by deriving the fundamental signal-to-noise ratio (SNR) of the measurement process, in standard deviations of the field amplitude. The sensitivity is then defined as the incident signal field amplitude ($\Erf$) \footnote{There is an extra factor of $\sqrt{2}$ that arises depending on if the measurement is of the rms or peak field amplitude. If the carrier frequency is within the instantaneous measurement bandwidth, then the sensor is capable of directly reading out the peak electric field, otherwise the root-mean-squared (rms) value is typically output. In Figure \ref{fig:PeakSensitivity} all fields are referenced to their peak value even if the underlying measurement is of the rms amplitude.} that results in $\SNR=1$ as measured in a one second integration time. If the $\SNR\propto\Erf$ and the noise is white, this definition is equivalent to the standard field sensitivity unit of \si{\Vm\per\sqrthz} and can be straightforwardly scaled to other measurement times and field amplitudes. However, this is not universally true for the Rydberg sensor, where $\SNR\propto\Erf^\beta$ with $\beta$ ranging between 1 and 2 (as discussed in Section \ref{subsec:SpecModel}), so we use our more general definition of sensitivity to avoid misinterpretation. Note that often, especially outside of a laboratory context, environmental noise dominates the overall noise profile of the measurement result. Here we choose to set aside these external noise sources in order to characterize the basic sensor technologies themselves. When evaluating the sensitivity requirement for an E-field sensor in a particular application, these external noise sources are important to consider.

\begin{figure}
    \centering
    \includegraphics[width=\linewidth]{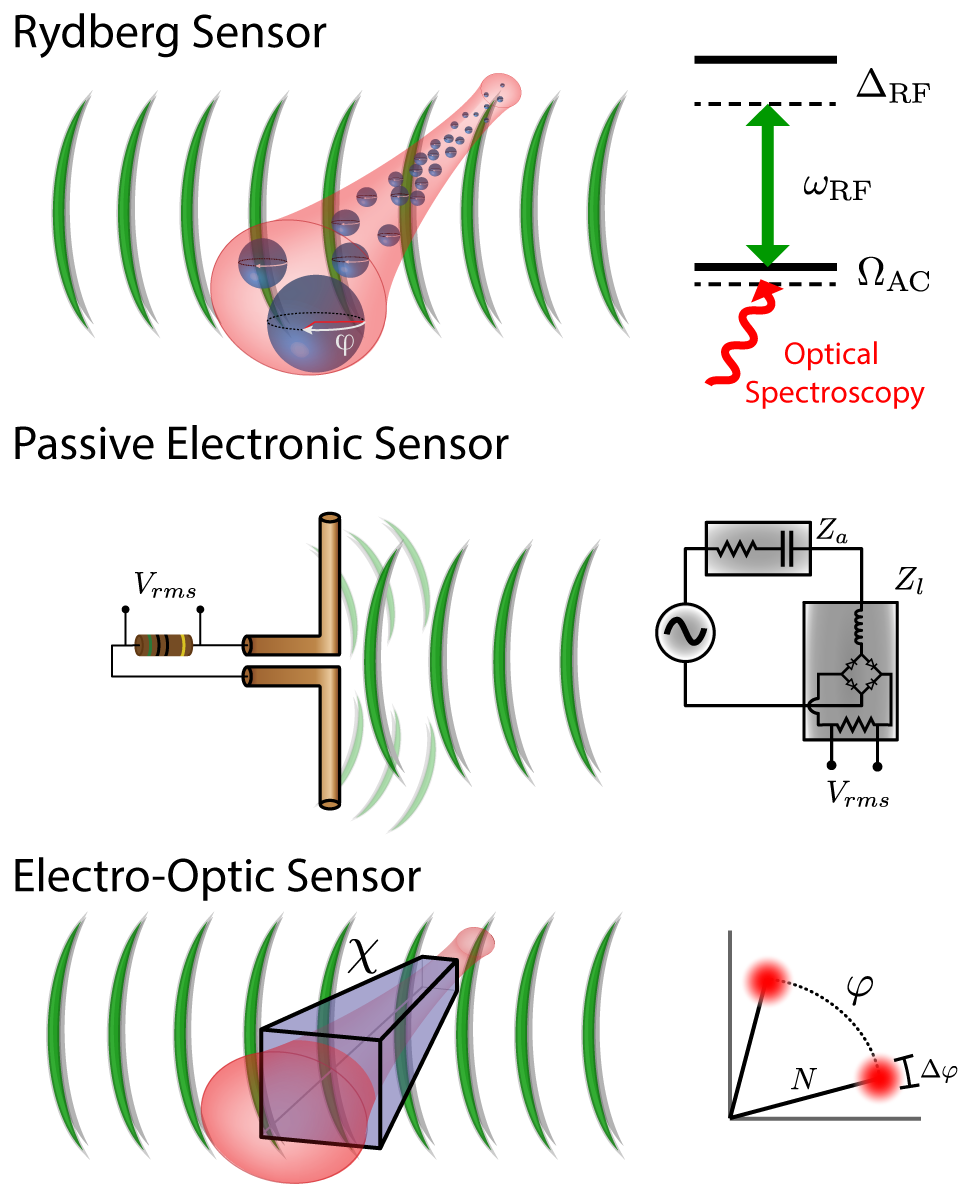}
    \caption{Examples of electric field sensors: 
    \textbf{Rydberg Sensor)} A dilute vapor of highly excited Rydberg atoms are perturbed by an incident RF field. These perturbations shift the atomic energy levels and are detected using optical spectroscopy.
    \textbf{Passive Electronic Sensor)} The incident RF field is coupled to a passive sensing load using a center-fed dipole antenna. The strength of this coupling depends on the dimensions of the antenna relative to the field wavelength and the impedance matching to the load.
    \textbf{Electro-Optic Sensor)} The incident RF field induces a change in the refractive indices of the crystal. A probing optical field is then used to measure this change, typically using a Mach-Zehnder interferometer.}
    \label{fig:Background}
\end{figure}

Figure \ref{fig:Background} illustrates the three primary electric field sensors discussed in this work: the Rydberg sensor and two comparison sensors based on dipole-coupled passive electronics or electro-optic crystals. This figure also diagrams a simple, conceptual model that governs each underlying sensor: a two-level atom, an equivalent circuit, and a phasor diagram, respectively. 

Atoms are best described in the language of quantum mechanics, and Rydberg sensors can rightfully be considered ``quantum sensors'', particularly as they have performed at the standard quantum (shot noise) limit \cite{facon_sensitive_2016,cox_quantum-limited_2018}. Their sensitivity to electric fields relies on large electric dipole moments and the corresponding energy shifts to the atomic spectroscopy that are detected optically \cite{osterwalder_using_1999}. Although not essential, Rydberg sensors to date have generally relied on the EIT method for state preparation and sensor readout \cite{sedlacek_microwave_2012}. The EIT dark state, which is a coherent quantum superposition of ground and Rydberg states, results in a narrow spectral resonance well suited for precision measurement. In a broader context Rydberg atoms have been used to create exotic quantum entangled states \cite{brune_manipulation_1992}, and shown promise in the field of quantum information science \cite{saffman_quantum_2010,graham_rydberg_2019}. Though quantum properties are not the primary focus of the present work, it is worth highlighting that quantum sensors bring important general features such as the ability to achieve sub-shot noise level measurements.

Electro-optic (EO) crystal-based sensors, in which changes of the indices of refraction due to the presence of electric-fields are detected with lasers, are similar in many ways to Rydberg sensors. Both are dielectric and can be made without any conductive material near the sensing volume. They are therefore transparent over a wide range of electric-field frequencies and this enables a non-destructive sensing interaction. Additionally, both devices work by transducing the RF information onto an optical field, lending to highly effective interferometric phase readout. The interaction strength between the field and sensing element can be characterized by the material's susceptibility, $\chi$. One difference between EO crystals and a Rydberg vapor is that crystals typically use a second order $\chi^{(2)}$ nonlinearity while Rydberg vapors rely on a third order $\chi^{(3)}$ due to the vapor's spatially centrosymmetric nature nullifying its $\chi^{(2)}$ response. 

Traditional electronics represent the most common and highly developed forms of electric field sensors due to their long history, low cost, and familiar implementation. In this work we restrict our consideration to a center-fed dipole antenna of length \SI{1}{\centi\meter}, similar in size to the Rydberg sensing volume, with electronic readout using an ideal rectifier and a load resistor. This system is readily modeled as a voltage divider connected to an ideal voltage source.  While simple, this model reasonably characterizes the nominal performance of room-temperature electronic readout, which is fundamentally limited by thermal noise.  It does not account for non-Foster circuit elements that can allow for higher sensitivities \cite{hansen_electically_2006}.    

We recognize that there is a wide array of electric field sensors that we do not consider in this work. Our motivation is to provide a foundation for broader consideration of the application space for Rydberg sensors in the context of some common sensor platforms rather than an exhaustive comparison with the entire field.

\section{Sensitivity Comparison in Low Frequency Regime}\label{sec:QuasiDC}

If we limit the frequency of the incident electric field to be much less than a device's lowest natural resonance, one can obtain simple analytic solutions for the sensitivity of Rydberg, passive electronic, and EO sensor systems. In the context of antenna engineering, this is known as the electrically small regime where $\ell/\lambda\ll1$, with $\ell$ being the physical size of the sensor, and leads to fundamental effects such as the Wheeler-Chu limit\cite{wheeler_fundamental_1947,chu_physical_1948}. In this section we derive analytic formulas for the sensitivity of a Rydberg sensor, a small dipole electronic sensor, and an electro-optic sensor in the low frequency regime.

\subsection{Rydberg atoms}
\label{QNoiseTheory}

The common method for implementing a Rydberg electric field sensor involves optical pumping to prepare the atoms into a sensitive Rydberg superposition state, interaction of that state with the incident electric field via Stark shifts, then optical readout of the collective phase shift of the initial state; see simplified diagram in Figure \ref{fig:Background} and more detailed diagram in Figure \ref{fig:setup}. As described in our previous work \cite{cox_quantum-limited_2018}, the SNR of this process is ultimately limited by the phase resolution of the readout stage due to the finite number of participating Rydberg atoms and the standard quantum limit. Here we outline that derivation. 

While the Rydberg sensor analysis and qualitative trends in this manuscript are transferable to other species of Rydberg sensor, the details and quantitative results will change depending on the specific atomic species used. We do not claim that our choice of species (rubidium) is inherently superior. Such a decision would depend on details of the intended use. For example, desires for specific RF resonances, laser colors, or vapor density/operating temperature will influence the choice of species.

We begin by defining the SNR as $\varphi/\Delta\varphi$ where $\varphi=\Omega\tau$ is the accumulated phase between two quantum states in an evolution time $\tau$ due to the atomic frequency shift $\Omega$. The phase noise $\Delta\varphi$ is assumed to be at the standard quantum limit, \emph{i.e.}, $\Delta \varphi_{SQL} = 1/\sqrt{N}$, with $N$ being the number of atoms.

The finite coherence time of our atomic sensor, $T_c$, gives an effective measurement/evolution time, $\tau$, that depends on whether the measurement time, $t$, is greater or less than $T_c$.
\begin{equation}\label{eq:tau}
    \tau = 
    \begin{cases}
        t & t< T_c \\
        \sqrt{\frac{T_c}{t}}t & t > T_c
   \end{cases}
\end{equation}
When $t>T_c$ an optically-pumped superposition state will, on average, collapse before readout and be repumped. This reset leads to a smaller observed phase shift from an ensemble of atoms by a factor of $\sqrt{T_c/t}$ \cite{kitching_atomic_2011}. The coherence time $T_c$ is influenced by many experimental details including transit broadening from the thermal motion of the atoms. In this work we assume a conservative $T_c=\SI{52}{\nano\second}$.

For electric field frequencies much lower than any atomic resonance considered (\emph{i.e.} which is $\SI {\sim 2}{\giga\hertz}$ considering the $n=100$ D state of Rb), the frequency shift due to the incident field can be estimated using the DC Stark shift \cite{gallagher_rydberg_2005},
\begin{equation}\label{eq:alphaEst}
    \Omega=-\frac{1}{2}\alpha E^2\approx-\frac{1}{2}\left( \frac{a_0 e}{\hbar}\right) ^2 \frac{\hbar n^7}{R_\infty}E^2
\end{equation}
where $a_0$ is the Bohr radius, $\hbar$ is the reduced Planck's constant, $e$ is the charge of the election, $R_\infty$ is the Rydberg constant and $n$ is the principal quantum number of the Rydberg state. The polarizability $\alpha$ of the Rydberg state can be approximated as shown under the rotating wave approximation. Finding the field, $\Erf_\text{Rydberg}$, which makes the SNR equal to one yields
\begin{equation}\label{eq:RydSens}
    \Erf_\text{Rydberg}=N^{-\frac{1}{4}}\sqrt{\frac{2}{\alpha \tau}}\approx N^{-\frac{1}{4}}\sqrt{\frac{2\hbar R_\infty}{a_0^2 e^2 n^7 \tau}}
\end{equation}

We see that the strength of a Rydberg sensor, in terms of sensitivity, lies in the scaling of the polarizibility with principle quantum number, $n$, and the potential to use many identical atoms, $N$, which can be packed within one electric field wavelength thanks to their small relative size.  However, because the SNR scales with $\Erf^2$, the sensitivity's scaling with $n$ and $N$ is suppressed by the additional square root. For example, to reduce $\Erf_\text{Rydberg}$ by a factor of 10 would require a factor of $10^4$ more atoms. 

The accuracy of the approximation of $\alpha$ in Eq.~\ref{eq:alphaEst} depends on how many nearby atomic resonances are taken into account and the validity of the rotating wave approximation. For example, estimating the polarizability due to a low frequency field considering only the next nearest state from $\state{100}{D}{5/2}$ yields \SI{-45.4}{\giga\hertz\cmV\squared}, whereas accounting for all nearby Rydberg states (calculated numerically \footnote{Calculated using the \texttt{ARC-Alkali-Rydberg-Calculator} Python package.}) yields \SI{-8.6}{\giga\hertz\cmV\squared}. The impact of each subsequent state diminishes as the respective detunings get larger, but in our particular case the second nearest state plays a significant role since $D$ states sit rather symmetrically between $P$ and $F$ states, meaning that the second nearest state is almost equally detuned as the first. It happens that the second state contributes a counteracting shift, which reduces the effective electric field sensitivity for the given target state (as in the example given). In Figure \ref{fig:PeakSensitivity}a we account for all states and plot the low frequency Rydberg sensor sensitivity using the numerically obtained polarizability with atom numbers $N=10^3$ and $N=10^4$ shown as solid and dashed red traces respectively. These numbers represent optimistic values for a Rydberg sensor using EIT readout where velocity selective probing significantly reduces the number of participating atoms \cite{urvoy_optical_2013,fan_effect_2015}. High Rydberg atom densities can also lead to complicating ion formation and Rydberg-Rydberg interactions.

While the quadratic signal scaling is a disadvantage when sensing weak fields directly, it opens the possibility of superposing a known strong field, $E_\text{bias}\gg \Erf$, to amplify the effect of the weak field under test (\emph{i.e.} heterodyning). Under this assumption the sensor's SNR scales linearly with $\Erf$. Assuming the uncertainty and noise of $E_\text{bias}$ is less than the SQL, this technique improves the minimum detectable field to
\begin{equation}
    \Erf_\text{Ryd-bias}\approx\frac{1}{\alpha\tau\sqrt{N}E_\text{bias}}
\end{equation}
Using this method, a sensitivity better than \SI{1}{\mVm\per\sqrthz} has been recently observed using $n=100$ Rydberg atoms and a non-zero $E_\text{bias}$ for $\orf<\SI{10}{\kilo\hertz}$ \cite{jau_vapor-cell-based_2019}.

\subsection{Passive Electronics}
\label{subsec:QuasiDC-Dipole}

As size constraints generally affect the performance of any sensor, we consider a short dipole antenna that is comparable in size to our Rydberg vapor sensing volume and connected to passive readout electronics. For low frequencies this means that the antenna will be electrically small (\emph{i.e.} $\lambda\gg\ell/10$). Along with loop antennas, dipole antennas form the majority of electrically small antennas in use today \cite{stutzman_antenna_2012}. 

To determine this sensor's fundamental SNR, we estimate the signal strength by modeling the short dipole antenna as an ideal voltage source, with an intrinsic impedance dependent on the geometry, coupled to a read out resistor through an ideal full-wave rectifier (see Fig.~\ref{fig:Background}). We assume the dominant noise source to be the sense resistor's thermal rms Johnson noise, $\sqrt{4k_b T R\Delta f}$. Here $k_B$ is Boltzmann's constant, $T=\SI{300}{\kelvin}$ is room temperature, $R$ is resistance, and $\Delta f=1/t$ is the measurement bandwidth.

The magnitude of the voltage source signal is given by the product of the electric field and the full length, $\ell$, of the dipole. The impedance of the antenna, $Z_a$, is predominantly capacitive and is given to good approximation as \cite{hansen_electically_2006}: 
\begin{equation}\label{eq:ESA-SNR}
    Z_a \approx i\left[\frac{Z_0}{\pi}\left(1-\ln{\left(\frac{\ell}{2a}\right)}\right)\cot{\left(\frac{\orf \ell}{2 c}\right)}\right]
\end{equation}
where $a$ is the radius of the conductor, $\orf$ is the angular frequency of the incident radiation, $Z_0$ is the impedance of free space, and $c$ is the speed of light. The signal strength will depend on the degree of impedance matching between the antenna and load resistor, $R_l$. The equivalent circuit model reduces to that of a simple voltage divider, and the SNR of the measurement is
\begin{equation}
    \SNR_\text{Dipole}=\frac{\Erf\ell}{\sqrt{4k_b T R_l\Delta f}}\frac{R_l}{|Z_l+Z_a|}
\end{equation}
where $Z_l$ is the lumped impedance of the load including the readout resistor and any matching network.

If no matching network is used (\emph{i.e.} $Z_l = R_l$) the SNR is maximized at a particular frequency by matching $R_l=|Z_a|$. The associated rms minimum detectable field in a one second measurement is
\begin{equation}
    \Erf_{\text{Dipole}}=\sqrt{\frac{8k_B T|Z_a|}{\ell^2}}
\end{equation}
This result, with $\ell=\SI{1}{\centi\meter}$, $a=\SI{300}{\micro\meter}$, and optimized $R_l=\SI{2.1}{\mega\ohm}$ at \SI{1}{\mega\hertz}, is shown in Figure \ref{fig:PeakSensitivity}a as the solid black trace. While not flat across this portion of the spectrum, the sensitivity is significantly greater than the Rydberg sensor. This is to be expected since the dipole antenna, even in this regime, acts as a superior coupler of the incident field than the free-space atoms. Using the dipole coupler with the Rydberg sensor would lead to significantly higher sensitivity as well. 

Enhanced sensitivity at a desired frequency can be accomplished, at the cost of sacrificing carrier spectral range and instantaneous bandwidth, by the addition of an impedance matching network in the form of an inductor that cancels the capacitance of the antenna to create a resonant dipole with higher $Q$-factor. An example of this with $R_l=\SI{50}{\ohm}$ is shown in Figure \ref{fig:PeakSensitivity}a as the dashed black trace. This line approaches the Chu-Wheeler limit for the \SI{1}{\centi\meter} electrically small antenna, and we indeed see higher sensitivity, though only over a very small bandwidth.

In this model we have assumed an ideal, passive rectifier, since rectification is necessary in order to measure a non-zero rms voltage over the sense resistor. In practice this is implemented using diodes. For small signal inputs (which are implicit when defining minimum detectable field) this means the circuit is driving a nonlinear load with non-zero forward voltage drop \cite{kanda_analytical_1980,ladbury_electrically_2002}. This presents a significant technical limitation to the realizable minimum field for passive electronic readout, on the order of \SI{1}{\Vm\per\sqrthz}, that we have not included in our model \cite{kanda_isotropic_1987}. This issue can be avoided using active components/measurement techniques such as transistors or RF heterodyning. While we do not explicitly consider these detection schemes, both are ultimately limited by Johnson noise and would therefore have similar idealized performance to the simple model presented here.

\subsection{Electro-Optic Crystals}
\label{subsec:EO}

The Pockels effect in an electro-optic crystal is an established mechanism for detecting electric fields \cite{powers_fundamentals_2011,duvillaret_electro-optic_2002}. In a similar way to the Rydberg sensor, we define the signal from a Pockels-based EO sensor to be the optical phase shift on a probing field due to the RF field in the crystal medium. Measurement of this phase is typically done using a Mach-Zehnder interferometer configuration. The noise limit in this case is determined by optical shot noise.

Assuming proper polarization when entering a $\overline{4}3m$ or $23$ crystal, the relative phase shift on the probe light is
\begin{equation}\label{eq:Waveplate}
    \varphi = \frac{2\pi L}{\lambda_0} \Delta n= \frac{4\pi L}{\lambda_0} \frac{n_0^3 r \Erf}{1+\sqrt{\epsilon_r}}
\end{equation}
where $L$ is the length of the crystal (interaction length), $\lambda_0$ is the wavelength of the probe in vacuum, $\Delta n$ is the difference in the index of refraction for the ordinary and extraordinary axes of the material, $n_0$ is the index of refraction of the ordinary axis, $r$ is the EO coefficient, and $\epsilon_r$ is the dielectric constant of the bulk crystal. The factor of $1+\sqrt{\epsilon_r}$ accounts for the reduction of free-space $\Erf$ inside the crystal due to its dielectric constant \cite{wu_ultrafast_1996}. Various choices of crystals exist, as discussed in References  \cite{wu_ultrafast_1996,duvillaret_electro-optic_2002}. For the sake of comparison, we have chosen ZnTe with $r\approx\SI{4.0}{\pico\meter\per\volt}$, $\epsilon_r=10.1$, and $n_0\approx 2.8$ at a probing wavelength $\lambda_0=\SI{633}{\nano\meter}$.

The phase uncertainty due to photon shot noise is $\Delta\varphi=1/2\sqrt{\overline{N}(t)}$, where $\overline{N}(t)$ is the average number of photons from the probe light expected in a measurement time $t$. Again taking the minimum detectable E-field, $\Erf_{\text{EOM}}$, to be when $\SNR=\varphi/\Delta \varphi=1$, we find 
\begin{equation}
    \Erf_{\text{EOM}} = \frac{\lambda_0(1+\sqrt{\epsilon_r})}{8\pi n_0^3 r L \sqrt{\overline{N}(t)}}
\end{equation}

This result is shown as the green trace of Figure \ref{fig:PeakSensitivity}a, taking $L=\SI{1}{\centi\meter}$ and probe power $P_\text{probe}=\SI{150}{\micro\watt}$, or $\overline{N}(\SI{1}{\second})=(P_\text{probe}t)/(h f_\text{Probe})=4.8\times10^{14}$ (where $h$ is Planck's constant and $f_\text{Probe}$ is the frequency of the probe light). Since the SNR is linearly proportional to $\Erf$, there is a more favorable scaling with photon number as compared to the scaling with atom number in the Rydberg sensor. Because of this, the EO sensor performs at a similar level to the Rydberg sensor despite comparatively weak nonlinear susceptibility. While the sensitivity also scales favorably with the crystal length and probe power, it is not practical to arbitrarily increase both due to the challenges of large crystal growth and achieving optimal, shot-noise limited Mach-Zehnder performance for ever increasing photon number. Demonstrated performance of an EO sensor on the order of \SI{1}{\mVm\per\sqrthz} has been reported in the literature \cite{toney_detection_2014}.

Another point of comparison is the minimal perturbation to the measured field from the dielectric EO crystal. Similar to the Rydberg sensor, the EO sensor head can be made without conductors, enabling a relatively non-destructive measurement. The remaining perturbation to the field is due to the step in index of refraction at the crystal surface, which can be significant for EO crystals \cite{wu_ultrafast_1996}. Comparing with the Rydberg sensor, the Rydberg vapor presents a significantly smaller index change and correspondingly smaller perturbation. However, the glass cell containing the vapor often presents a significant index change and must also be considered.

Finally, the sensitivity is relatively flat and independent of $\orf$, which is convenient for sensor operation. Resonances that limit this flat response do arise, particularly as the RF wavelength approaches the length scale $\ell$, and these depend strongly on the mechanical design of the sensor. To reflect these considerations, we have extended the low-frequency result into Fig.~\ref{fig:PeakSensitivity}b up to \SI{\sim20}{\giga\hertz}, an operational range that commercial EO sensors readily achieve.

\section{Wide Spectrum Sensitivity of the Rydberg Sensor}
\label{sec:Spectrum}

In this section we extend the quantitative measure of the Rydberg sensor's sensitivity to cover a wider carrier spectral range.
At frequencies \SI{>100}{\mega\hertz}, atoms excited to a Rydberg state provide a structured spectrum of sensitivity to electric fields due to strong resonant and off-resonant interactions with many dipole-allowed transitions to nearby Rydberg states. As discussed above, these interactions produce Stark shifts with respect to the target Rydberg state that can be detected via optical spectroscopy. 
The scaling of this shift with the applied electric field amplitude, $\Erf$, depends on the frequency of the radiation, $\orf$, relative to the atomic resonances.
Near resonance, the Stark shift takes the form of Autler-Townes splitting (a special case of the AC Stark effect) and is proportional to $n^2\Erf$. Off-resonance, the shift takes the form of a general AC Stark shift and is proportional to $n^7 \Erf^2$. Using both regimes, sensitivity to fields with frequencies ranging from \SI{10}{\kilo\hertz}\cite{mohapatra_giant_2008,cox_quantum-limited_2018} to \SI{1}{\tera\hertz}\cite{wade_real-time_2017} have already been demonstrated. 

Here we develop a theoretical treatment based on semi-classical Floquet theory to estimate the minimum detectable field of the Rydberg sensor for arbitrary carrier frequencies that is valid for sub-ionizing field strengths. 
We also confirm the theoretical model via comparison with experimental data obtained using a commercially-available wideband antenna operating over \SIrange{1}{20}{\giga\hertz} for three particular Rydberg states. 

\subsection{Modeling}
\label{subsec:SpecModel}

If we first limit consideration to relatively weak field strengths common in communication or remote sensing applications and frequencies near atomic resonances, a standard textbook model of the AC Stark shift using the Rotating Wave Approximation (RWA) is valid. 
This model assumes a two-level system with a strong coupling field detuned much less than the transition frequency between the two levels. 
Going to the rotating frame of the RF field and ignoring the counter-rotating term, the AC Stark shifted energies of the two levels take the form of
\begin{equation}
    \Omega_\text{AC}=\frac{1}{2}\left(\Delta\pm\sqrt{\Delta^2+\Omega^2}\right)
\end{equation}
where $\Delta$ is the detuning of the incident RF field from resonance, $\Omega=d\cdot\Erf/\hbar$ is the resonant Rabi frequency of the RF field and $d$ is the dipole moment of the transition. 
Which shift corresponds to the lower energy state depends on the sign of $\Delta$: if $\Delta>0$ corresponding to a blue detuning the minus sign is used, $\Delta<0$ uses the positive sign. 
At $\Delta=0$, both roots have the same magnitude ($\Omega/2$) resulting in the common-mode splitting known as Autler-Townes splitting.

\begin{figure}[tb]
    \centering
    \includegraphics[width=\columnwidth]{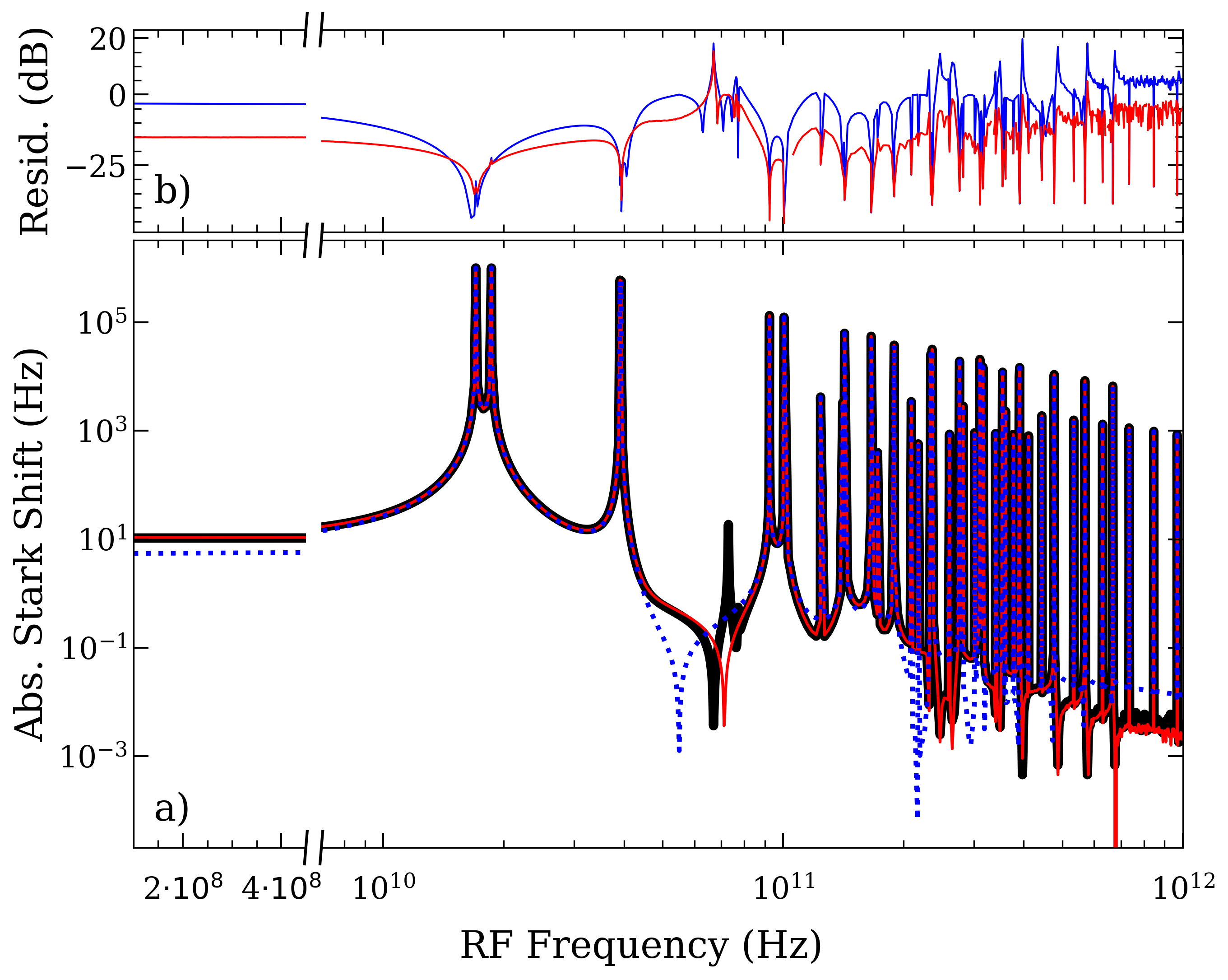}
    \caption{Comparison of Rydberg Models: a) The black line is the full Floquet model calculated for $\state{50}{D}{5/2}[1/2]$ state with $\Erf=\SI{100}{\milli\volt\per\meter}$. The red line is the reduced Floquet model. The blue line is the perturbative model. b) The normalized residuals between the two approximate models and the full Floquet theory are shown above in \si{\decibel}; note \SI{0}{\decibel} represents an error of \SI{100}{\percent}.}
    \label{fig:ModelComparison}
\end{figure}

The total Stark shift from multiple nearby levels is found by summing together the contribution of each two-level system calculated separately. 
Figure \ref{fig:ModelComparison} shows this model's estimate of the absolute Stark shift of the $\state{50}{D}{5/2}[1/2]$ Rydberg state due to a $\Erf=\SI{100}{\milli\volt\per\meter}$ field versus frequency in comparison with the more complete Floquet models developed below. Near atomic resonances this simple model has good agreement. Further from resonances where the detuning is on order with the transition frequency the counter-rotating term cannot be ignored and the validity of the approximation breaks down. While less accurate in these far-detuned regimes, this model is very fast to calculate numerically compared with the Floquet model and therefore can be useful for rough sensitivity estimates.

A more complete model is derived using semi-classical Floquet theory, outlined in detail in Ref. \cite{chu_recent_1985}. Floquet theory is capable of modeling Stark shifts for arbitrary field amplitude and frequency, and represents a more complete solution when determining the Rydberg sensor's sensitivity \cite{anderson_optical_2016}. Here we briefly outline the Stark shift calculation procedure using this theory.

We start with the time-dependent Schr\"odinger equation
\begin{equation}
    \left[H_0+V(t)\right]\Psi(t)=i\hbar\frac{\partial}{\partial t}\Psi(t)\label{eq:Shroedinger}
\end{equation}
where perturbation potential $V$ from the RF field is periodic in time such that $V(t+\tau)=V(t)$ ($\orf\tau=2\pi$) and the bare atomic Hamiltonian $H_0$ has eigenfunctions such that $H_0\ket{\alpha}=E^0_\alpha\ket{\alpha}$,  $\braket{\beta|\alpha}=\delta_{\alpha\beta}$. The precise numerical values for the energy levels ($H_0$) and dipole moments ($V$) are found using numerical integration as provided by the Alkali-Rydberg-Calculator (ARC) Python package \cite{sibalic_arc_2017}.

The Floquet theorem states that the periodic nature of the perturbation potential implies the solutions to the Schr\"odinger equation should also be periodic such that $\Psi(t)=e^{-i\bm{\epsilon} t/\hbar}\Phi(t)$ where $\bm{\epsilon}$, known as the quasi-energies, is a diagonal matrix of unique, real numbers, $\epsilon_\beta$, up to integer multiples of $2\pi/\tau$ and $\Phi(t+\tau)=\Phi(t)$ is a matrix of corresponding periodic functions. The time-periodic Hamiltonian $H(t)=H_0+V(t)$ and $\Psi(t)$ can be expanded into the Floquet-state basis $\ket{\alpha n}=\ket{\alpha}\ket{n}$ where $\ket{n}$ are Fourier vectors corresponding to harmonics of $\orf$ such that $\braket{t|n}=e^{in\orf t}$.
\begin{align}
    \bra{\alpha}\Psi(t)\ket{\beta}&=\sum^\infty_{n=-\infty}\Phi^{(n)}_{\alpha\beta}e^{in\orf t}e^{-i\epsilon_\beta t/\hbar}\label{eq:Phi}\\
    \bra{\alpha}H(t)\ket{\beta}&=\sum^\infty_{n=-\infty}H^{[n]}_{\alpha\beta}e^{in\orf t}\label{eq:H}
\end{align}
Here $H_{\alpha\beta}^{[n]}$ represents the $n$th-order component of the Fourier expansion of the Hamiltonian (\emph{i.e.} $H^{[0]}=H_0$ and $H^{[\pm1]}=V$).

Solving for $\bm{\epsilon}$ and $\Phi(t)$ then gives the energy shifts to the bare atomic states and the transition probabilities between states. These solutions typically must be found numerically since the large dipole moments between the many nearby Rydberg states lead to significant, competing interactions that all must be taken into account. A typical basis for $H_0$ includes Rydberg states with $n\pm10$ relative to the target state and orbital quantum number $\ell\leq20$ resulting in over 800 states. Moreover, multi-photon resonances are possible for relatively small applied fields meaning the infinite sums of Eqs.~\ref{eq:Phi} \& \ref{eq:H} can only be truncated to $n\simeq10$, each order adding a multiple of two of the nominal atomic basis to the Floquet basis.

Semi-classical Floquet theory has already been applied in the context of Rydberg electrometers for large field amplitudes \cite{anderson_optical_2016,miller_radio-frequency-modulated_2016,paradis_atomic_2019}, where the solve takes the form of numerically integrating the the time-evolution operator $U(t+\tau,t)$. Here we are not interested in the transition probabilities and can thus choose to use the simpler Shirley's time-independent Floquet Hamiltonian method \cite{shirley_solution_1965}. This is done by substituting Eqs.~\ref{eq:Phi} \& \ref{eq:H} into Eq.~\ref{eq:Shroedinger} to obtain an infinite dimension eigenvalue equation
\begin{equation}
    \sum_{\gamma m}\braket{\alpha n|H_F|\gamma m}\Phi^{(m)}_{\gamma \beta}=\epsilon_\beta\Phi^{(n)}_{\alpha\beta}
\end{equation}
where $H_F$ is a block tri-diagonal matrix with elements
\begin{equation}
    \braket{\alpha n|H_F|\beta m} = H^{[n-m]}_{\alpha\beta}+n\hbar\orf\delta_{\alpha\beta}\delta_{nm}
\end{equation}
In this case, because we are focused on weak fields, we can truncate $H_F$ to $n=\pm1$ while also avoiding the integration of the time-evolution operator. Furthermore, we can reduce the basis of $H_F$ to only include those Rydberg states that have direct dipole-allowed transitions with the target state. This reduces the basis from $\sim800$ to $\sim40$, significantly improving the speed of computation. We will refer to this reduced basis solution as the reduced Floquet model.

In Figure \ref{fig:ModelComparison} we choose an applied field of \SI{100}{\milli\volt\per\meter} and a single target state, $\state{50}{D}{5/2}[1/2]$, and show comparisons of predicted Stark shifts for the full Floquet model (black trace), the reduced Floquet model (red trace), and the RWA model (blue trace). The magnitude of the normalized residuals between the two Floquet models (shown red in part b) is mostly less than \SI{-10}{\decibel}, except for small regions around intermediate detunings where the differences are somewhat larger. For example, this discrepancy is most visible for the set of features around \SIrange{6}{9e10}{\hertz}, where the shifts from nearby states conspire to significantly suppress the response. The RWA model, where the Stark shift from each dipole-allowed transition is added together to produce an average shift of the target state, shows larger discrepancies with the full Floquet model except on atomic resonances where agreement is quite good. It may seem surprising that this model is as effective as it is (within a factor 2 for most of the frequency range), but this is due to the relative weakness of the applied RF field, which reduces the influence of the far-detuned resonances that violate the RWA assumptions.

\begin{figure*}[t] 
    \centering
    \includegraphics[width = \textwidth]{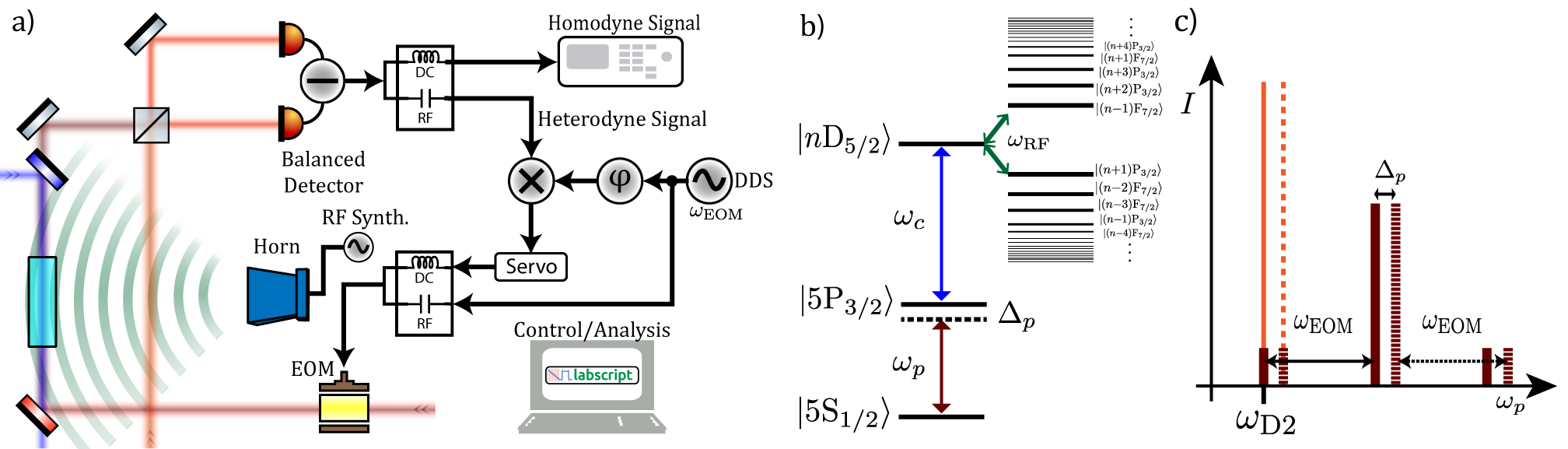}
    \caption{Experimental setup: 
    a) Optical and electronic configuration for the homodyne/heterodyne measurement. Experimental control and analysis is done using the \texttt{labscript} suite \cite{starkey_scripted_2013}. 
    b) Rubidium level diagram showing Rydberg excitation path and RF coupling to manifold of nearby Rydberg states.
    c) Homodyne/Heterodyne frequency spectrum. The solid red lines show the probe spectrum with EOM modulation applied. The solid orange line shows the frequency of the local oscillator (LO) reference. The laser frequency is set such that the LO and lower sideband of the probe are resonant with the $D2$ probing transition. Detuning the laser (dashed lines) moves both spectra in unison.}
    \label{fig:setup}
\end{figure*}

Each peak in Figure \ref{fig:ModelComparison}a is actually a pair of two nearby resonances (the lowest couplet near \SI{17}{\giga\hertz}, is visibly resolved) since the $\mathrm{D}_{5/2}$ states sit nearly symmetrically between the $\mathrm{P}_{3/2}$ and $\mathrm{F}_{7/2}$ states (illustrated in Fig.~\ref{fig:setup}b). Peaks at increasing RF frequency are couplets with increasing $\Delta n$, \emph{i.e.} 
$\state{50}{D}{5/2}\rightarrow\state{(50\pm \Delta n)}{P}{3/2}$ \&
$\state{(50\mp \Delta n)}{F}{7/2}$.

The structure of the frequency response shown in Figure \ref{fig:ModelComparison} has important implications for a wideband sensor. While the sensor has some measurable response at all frequencies, the discrete resonances (each $\mathsmaller{\lesssim}$\SI{10}{\mega\hertz} wide) provide amplified response at specific frequencies. This behavior is reminiscent of the harmonics of a dipole antenna (as seen in Figure \ref{fig:PeakSensitivity}b and described below), however the Rydberg sensor resonances are not related by harmonics. The implications are similar: the Rydberg sensor can preferentially detect many RF frequencies spread across its carrier spectral range without modification while effectively rejecting large portions where the atom response is significantly weaker. One important distinction is that the Rydberg atomic resonances are absolutely well known, and each atom is identical (a quantum advantage). Another distinction is that the Rydberg sensor signal depends primarily on the detuning of the RF field to the nearest resonance which does not convey the RF frequency directly. This can make determination of unknown RF frequencies more challenging and methods for addressing this will be the subject of future work.

While Figure \ref{fig:ModelComparison} shows the Stark shift of a single Rydberg target state over the full spectrum of considered frequencies, there are many Rydberg states that can be taken advantage of by simply tuning the Rydberg laser. Restricting the target Rydberg state to be $\state{n}{D}{5/2}[1/2]$ the optimal target state for maximizing sensitivity to a given RF frequency is shown above, in Figure \ref{fig:PeakSensitivity}b, calculated using the reduced Floquet model. A comparison to $\state{n}{P}{3/2}[1/2]$ and $\state{n}{S}{1/2}[1/2]$ target Rydberg states is located in the Supplemental Materials.

\begin{figure}[tb]
    \centering
    \includegraphics[width=0.85\columnwidth]{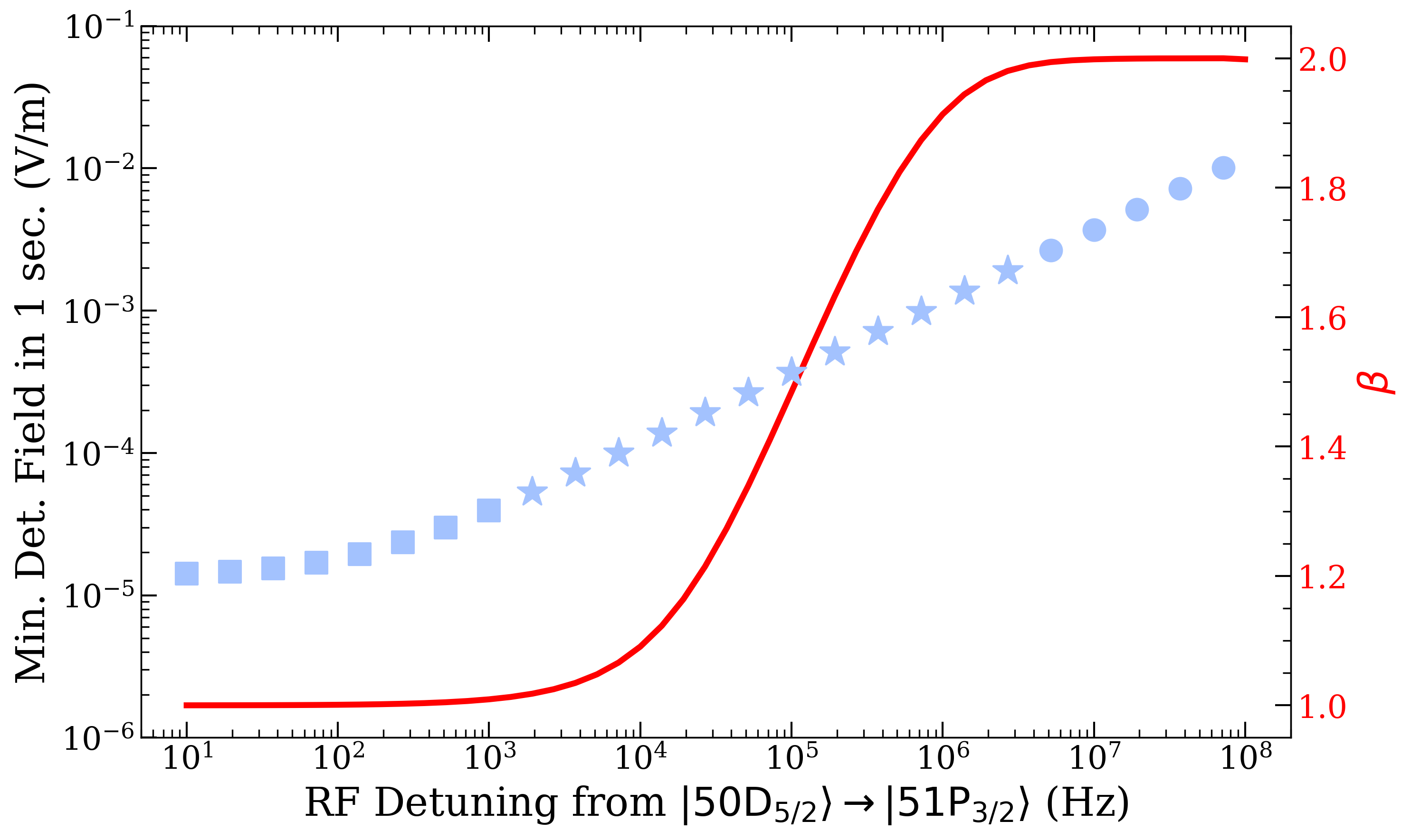}
    \caption{Scaling of minimum detectable field (blue) and $\beta$ (red) versus detuning from RF transition. The square, circle, and star symbols match those used in Fig. \ref{fig:PeakSensitivity}b and correspond to $\beta=1,2\,(\pm1\%)$ or somewhere in between, respectively.}
    \label{fig:SensBeta}
\end{figure}

To keep Figure \ref{fig:PeakSensitivity}b legible, we restricted the number of data points to include all dipole-allowed resonances as well as 300 more points distributed on a log scale per each decade of frequency.
Each point is calculated by first comparing the absolute Stark shift for a fixed $\mathcal{E}$ to identify the optimal $n$ for each frequency. We then use numerical optimization to find the $\SNR=1$ point for that optimal state. 

Numerical optimization is necessary because the scaling of the minimum detectable field is not known a-priori at every frequency, rather it depends on the detuning from nearby atomic resonances. As an example, Figure \ref{fig:SensBeta} shows how the sensitivity and the scaling of the SNR, $\beta$ (as in $\SNR\propto\Erf^\beta$), vary with detuning from the $\state{50}{D}{5/2}\rightarrow\state{51}{P}{3/2}$ transition. The precise width and transition point of the $\beta$ transition from 1 to 2 depends on the strength of the applied field, however the general shape is consistent for any Rydberg resonance. The corresponding value of $\beta$ for each point in Figure \ref{fig:PeakSensitivity}b is noted by the shape of the point: a square for $\beta=1$, a circle for $\beta=2$, and a star for an intermediate value. Knowing the value of $\beta$ allows one to use the results for any sensor in Figure \ref{fig:PeakSensitivity} to determine the SNR for any $\Erf$ in a 1 second measurement. 
For example, in a region where $\beta=1$, if $\Erf_\text{\SI{1}{\second}}=\SI{1}{\micro\volt\per\meter}$ then the expected SNR in standard deviation for a 1 second measurement of a \SI{100}{\micro\volt\per\meter} field of the same frequency is  $\left(\Erf/\Erf_\text{\SI{1}{\second}}\right)^\beta=100$. 
Scaling the minimum detectable field to other measurement bandwidths requires understanding of how the SNR scales with $t$. With the exception of the Rydberg sensor for $t<T_c$, this scaling takes the form of $\Erf_t=\Erf_\text{\SI{1}{\second}}t^{\sfrac{-1}{2\beta}}$, assuming white noise sources. If $\Erf_\text{\SI{1}{\second}}=\SI{10}{\micro\volt\per\meter}$ and $\beta=2$, the minimum detectable field in a $t=\SI{1}{\milli\second}$ measurement is \SI{56}{\micro\volt\per\meter}.

Figure \ref{fig:PeakSensitivity}b reveals a few basic patterns. First, the general rule for picking a target Rydberg level to use for a given field frequency is to choose the $n$ which allows the nearest to resonance. If there are multiple equally close resonances, use the $n$ with the smallest $\Delta n$ to the target level. Second, it is interesting to observe that there are two clusters of resonant transitions, one with smaller detectable field values and one with larger. The more sensitive set of transitions are made up of couplings to $\state{(n+\Delta n)}{P}{3/2}$ and $\state{(n-\Delta n)}{F}{7/2}$. The couplings with opposite sign are significantly weaker due to mismatch in the overlap of the radial wavefunctions with that of the target $\state{n}{D}{5/2}$ state, rendering them less sensitive \emph{i.e.} larger minimum detectable field values. For more details see the Supplemental Materials.

The choice to cut off consideration of Rydberg levels greater than $n=100$ is somewhat arbitrary, though such Rydberg levels have recently been used for low-frequency E-field measurements \cite{jau_vapor-cell-based_2019}. There are complicating factors not included in the model that cause concern as the Rydberg levels increase. These include the challenge of getting good EIT contrast and SNR, the need for more laser power to couple the same number of atoms, and various atomic interactions, particularly Rydberg-Rydberg interactions. 

For reference, current state of the art Rydberg sensor performance is \SI{\sim100}{\uVm\per\sqrthz} using $n=50$ \cite{meyer_digital_2018,kumar_rydberg-atom_2017}. Ideally this value would be near the bottom line of points in Fig.~\ref{fig:PeakSensitivity}b, but low quantum efficiency of detection, resulting in a reduced effective Rydberg atom number, has limited the minimum detectable field to date \cite{cox_quantum-limited_2018}.

\subsubsection*{Electrically Large Dipole Sensor}

As a point of reference, the gray line of Figure \ref{fig:PeakSensitivity}b shows the sensitivity of the same $\ell=\SI{1}{\centi\meter}$ center-fed dipole from Section~\ref{subsec:QuasiDC-Dipole}, but in an electrically-large regime ($\ell/\lambda\gtrsim1$) with a \SI{50}{\ohm} load sense resistor. This response is calculated using the induced-emf method to determine the intrinsic impedance and directivity gain of the dipole antenna as a function of frequency \cite{stutzman_antenna_2012,balanis_antenna_2005}. The sensitivity is then found using the same equivalent circuit model as the low-frequency passive electronic sensor, but with the length $\ell$ of Eq.~\ref{eq:ESA-SNR} replaced with an effective length, 
\begin{equation}
    \ell_\text{eff}=\sqrt{\frac{R_a \lambda^2 G(\lambda,\ell)}{\pi\eta Z_0}}
\end{equation}
where $G(\lambda,\ell)$ is the antenna gain and $\eta$ is the radiation efficiency. The resulting minimum detectable field is then given as
\begin{equation}\label{eq:FullDipole}
    \mathcal{E}_\mathrm{Dipole}=\frac{\sqrt{4k_b T R_l\Delta f}}{K}=\frac{\left|Z_a+Z_l\right|\sqrt{\pi\eta Z_0}\sqrt{4k_b T R_l\Delta f}}{R_l \sqrt{G(\lambda,\ell)R_a\lambda^2}}
\end{equation}
where $K$ is commonly known as the antenna factor. The half-wave frequency is \SI{\sim15}{\giga\hertz} and corresponds to the first high sensitivity dip. Higher frequency dips correspond to the $\ell=(m+1/2)\lambda$ frequencies. The low sensitivity peaks correspond to frequencies of integer wavelength multiples ($\ell=m\lambda$) where the center-feed point of the wire then sits at a node of the field, or in other words, becomes a high impedance point. We again note that this calculation is an idealized model. In practice it is difficult to accurately back out the absolute incident field strength due to the importance of parasitic elements at these frequencies.

Comparing with the Rydberg sensor, we observe that the passive electronics sensor has lower minimum detectable field at its optimal half-wave frequencies. However, coverage of the entire spectrum at this level may require difficult design and optimization of both the antenna and sensing system at each frequency. In contrast, the Rydberg system can be tuned to any of its optimal sensitivity points by simply tuning a laser frequency.

\subsection{Experiment}
\label{subsec:SpecExp}

We confirm the validity of our theoretical model by experimentally measuring the sensor response over a frequency range of \SIrange{1}{20}{\giga\hertz} using three different Rydberg target levels: $\state{50}{D}{5/2}$, $\state{60}{D}{5/2}$, and $\state{70}{D}{5/2}$. The frequency range of this measurement was dictated by the microwave source system; specifically the operational range of the widest-band transmission antenna readily available \footnote{A Schwarzbeck 9120D double-ridged broadband antenna. This and all other references to commercial devices do not constitute an endorsement by the U.S. Government or the Army Research Laboratory. They are provided in the interest of completeness and reproducibility.}.

The experimental setup and level diagram are shown in Figure \ref{fig:setup}a-b, and largely follow the standard Rydberg electrometer configuration: linearly polarized probe and coupling beams counter-propagate in a rubidium vapor cell establishing ladder Electromagnetically-Induced Transparency (EIT) spectroscopy of the Rydberg level, which is shifted by the presence of an RF field (for details see the Supplemental Materials). The transmitted probe light is measured using an optical homodyne method similar to those in Refs.~\cite{mohapatra_giant_2008,kumar_atom-based_2017} which allows for precise, photon-shot-noise limited measurements in both the phase and amplitude quadratures. Our implementation follows a modification used in Ref.~\cite{cox_generating_2015}. A single laser sent through acousto-optic modulators (AOMs) creates both the probe and reference beams; a subsequent electro-optic modulator (EOM) places sidebands on the probe beam such that the lower sideband is equal in frequency to the reference. An optical heterodyne signal between the carrier and the reference is used to actively stabilize the relative beam path phase, see Fig.~\ref{fig:setup}c. This method allows us to easily change measurements between amplitude and phase quadratures without the need for a second reference laser.

\begin{figure}[tb]
    \centering
    \includegraphics[width=\columnwidth]{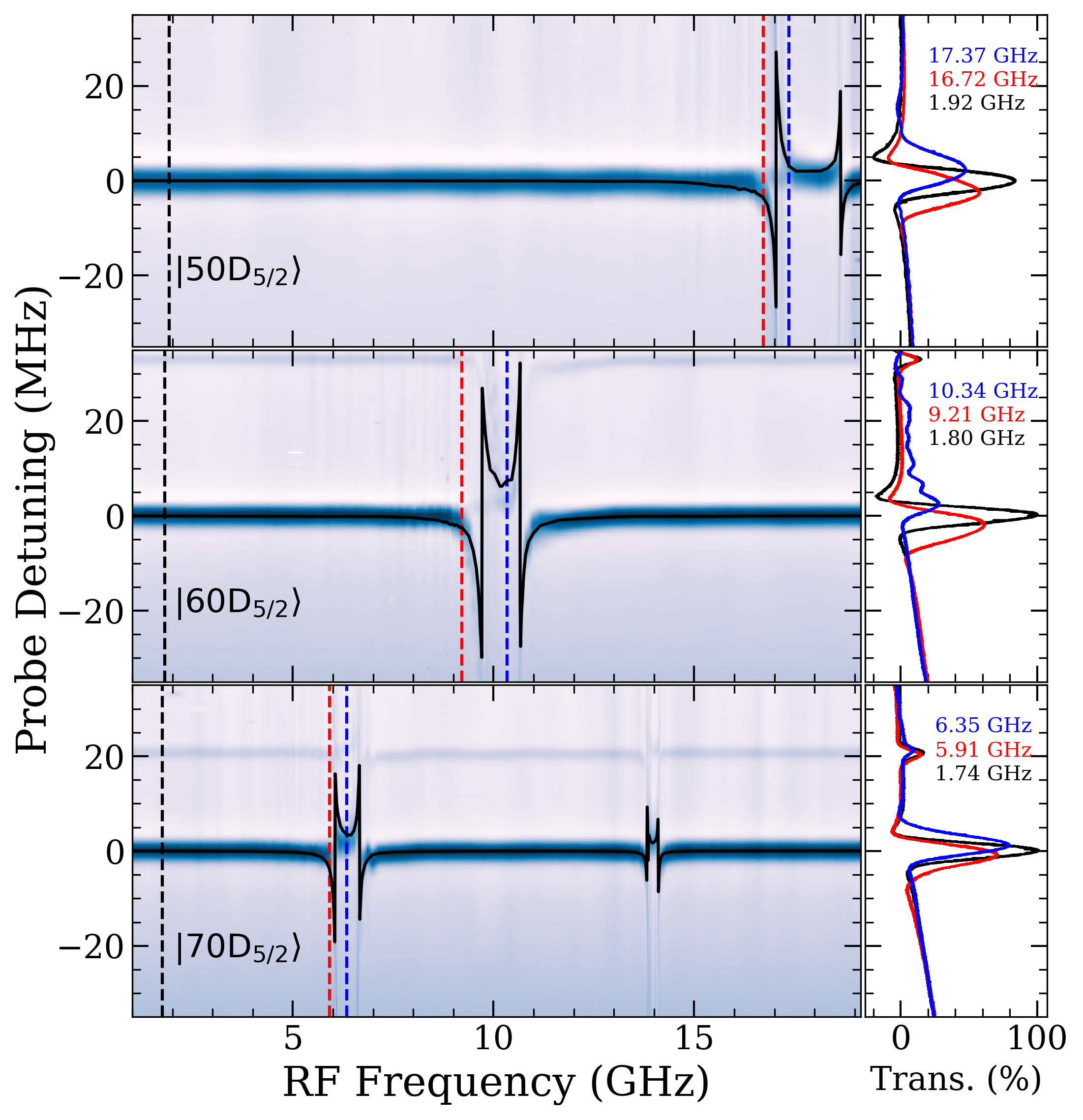}
    \caption{Atomic response versus RF frequency for $\state{n}{D}{5/2}$ target states. Ranging from top to bottom is $n=50,60,70$. The RF set power is \SIlist{16;12;3}{\dBm} respectively. The black lines show the expected level shifts from the Floquet theory. The vertical dashed lines indicate where which example sweep traces are shown to the right of each contour plot. The black dashed line shows the atomic response at far RF detuning, the red trace near the lowest couplet of resonances, and blue between the couplet resonances.}
    \label{fig:WidebandDataTheory}
\end{figure}

Figure \ref{fig:WidebandDataTheory} shows a contour plot of experimental data for \SIrange{1}{20}{\giga\hertz} fields and target Rydberg levels $n=50,60,70$, where the spectrum amplitudes are normalized to the bare EIT peak (\emph{i.e.} no RF present). In order to maintain similar-order Stark shifts for each level, the RF power was decreased with increasing $n$ ($P_\mathrm{set}=16, 9, \SI{3}{\dBm}$, respectively). To the right of each contour plot we show three slices for probe sweeps at RF frequencies that are far from resonance (black), near the lower resonant doublet (red), and inside the nearest resonant doublet (blue). One notices the red and cyan trace peaks are broadened relative to the far-detuned black traces, due to the influence of the various $m_J=1/2, 3/2, 5/2$ transitions. As the applied field strength is increased, these sublevels become resolved. The $n=60$ blue trace reveals some of this behavior as the nearby $\state{60}{D}{3/2}\rightarrow\state{61}{P}{1/2}$, which has slightly higher resonance frequency, experiences Stark shifts that overlap with those of the $\state{60}{D}{5/2}$ target state.

For direct comparison with Floquet theory, we calibrate the applied RF electric field amplitude through a resonant Autler-Townes splitting measurement of the $\state{50}{D}{5/2}\rightarrow\state{51}{P}{3/2}$ transition at \SI{17.0415}{\giga\hertz} \cite{sedlacek_microwave_2012,holloway_broadband_2014}. We then use the manufacturer-specified antenna gain profile and measured cable losses to extrapolate over the measured range, \SIrange{1}{20}{\giga\hertz}. The overlaid solid black lines show the Floquet-predicted shifts as a function of RF frequency, which shows good agreement with the measurements.

\begin{figure}[tb]
    \centering
    \includegraphics[width=0.95\columnwidth]{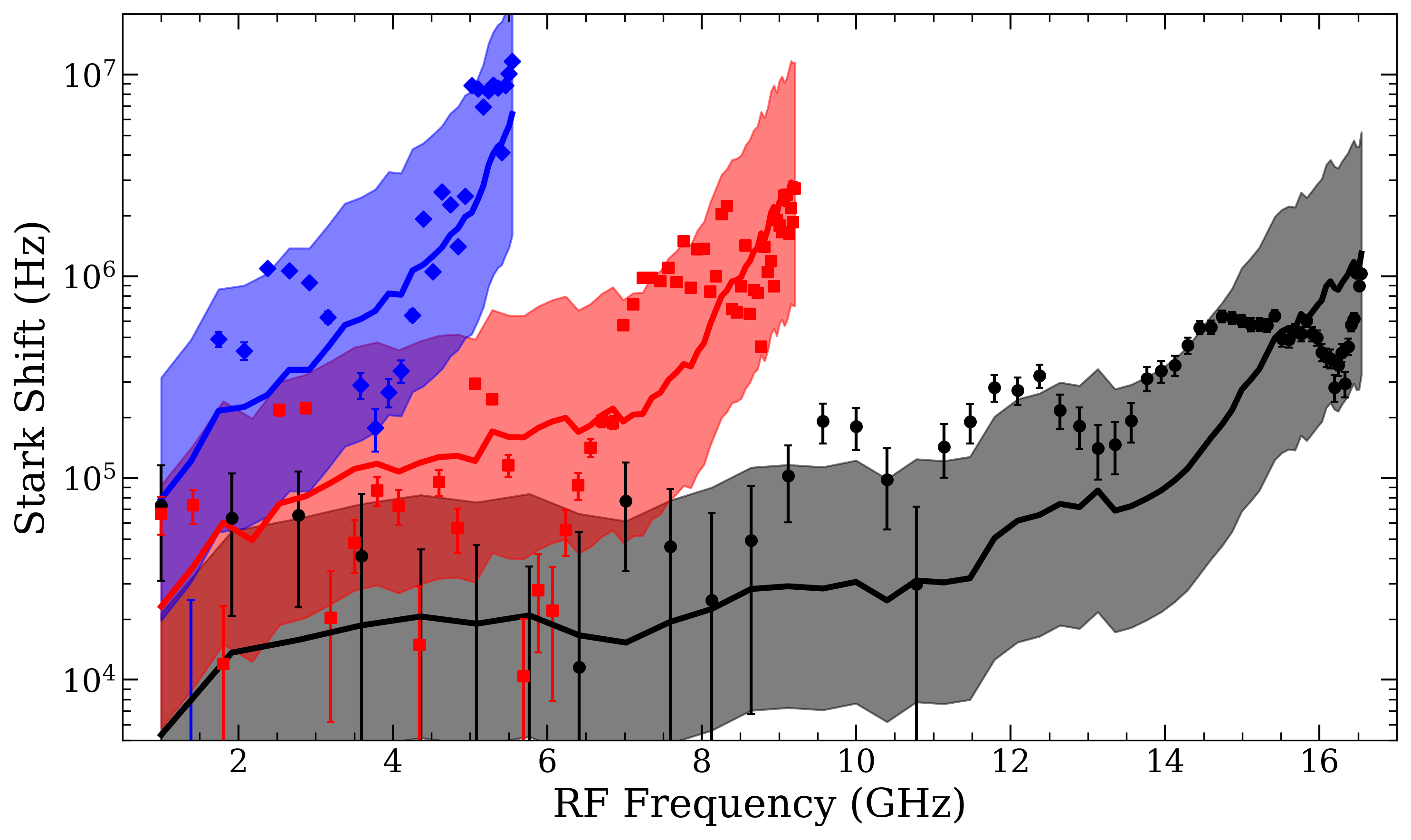}
    \caption{Atomic response versus RF frequency in the AC Stark regime. The black, red, and blue data correspond to $n=50, 60, 70$ $\state{n}{D}{5/2}$ target states, respectively. The lines represent the corresponding Floquet model predictions. The colored regions show the corresponding error in model estimate from field calibration error while the error bars show the corresponding error in experimental peak extraction.}
    \label{fig:ACRegimeDataTheory}
\end{figure}

We performed narrower probe sweeps with a fixed RF set power of $P_\mathrm{set}=\SI{16}{\dBm}$ for each level in order to make a more detailed comparison with theory. In Figure \ref{fig:ACRegimeDataTheory} we show the extracted Stark shift of the EIT peak relative to no applied field for $n=50,60,70$ as black circles, red squares, and blue diamonds, respectively. The error bars represent the sweep-to-sweep jitter of the measured EIT resonance. As expected, higher $n$ leads to larger Stark shifts for the same applied field, and as the frequency approaches a Rydberg resonance the Stark shift increases. The solid lines represent the Floquet predictions, and the shaded regions correspond to \SI{\pm3}{\decibel} changes in the applied RF field power, which accounts for fluctuations of environmental reflections/scatter, horn calibration error, and RF etalons within the vapor cell \cite{fan_effect_2015}. 
The difficulty in calibrating the wideband horn antenna versus the absolute atomic measurement uncertainty is demonstrated in this figure: while the accuracy of the Floquet predictions is limited by the horn calibration errors, the atomic measurements are significantly more accurate by nature and could be used to improve the calibration.
Note that the EIT resonance has a linewidth of $\Gamma\sim\SI{5}{\mega\hertz}$ and Stark shifts less than this width ($\Gamma/100$ or \SI{50}{\kilo\hertz}) are difficult to resolve accurately; this is particularly relevant for the $n=50$ data. Similar to the heterodyning in the low-frequency regime mentioned earlier, the addition of a biasing RF field can be helpful in addressing this challenge \cite{jing_quantum_2019,simons_rydberg_2019}.

These results reinforce confidence in the Floquet model as an effective predictor of Stark shifts due to arbitrary RF frequencies and amplitudes of interest. This allows us to not only determine optimal target Rydberg states for a given frequency and field, but also could enable the identification of unknown frequency fields by comparing Stark shifts on multiple target states.

\section{Conclusion}

With the current interest in Rydberg-based electric field sensors, there have been numerous creative proposals identifying potential application spaces. 
Rydberg sensors' wide spectral coverage and sensitivity have been touted as strengths, and are important figures of merit for many applications. 
We have presented multiple theoretical models of varying accuracy and computational complexity that predict the Rydberg sensor's spectral sensitivity over a wide range of field frequencies and amplitudes. We validated these models experimentally using a simultaneous homodyne/heterodyne measurement technique for three Rydberg levels over a frequency range of \SIrange{1}{20}{\giga\hertz}.

In this work we have also compared the Rydberg sensor to prominent, established electric field sensors; namely electro-optic crystals and dipole-coupled passive electronics. We used relatively simple models and assumed fundamental noise sources in order to be as general and broadly applicable as possible.
We find the Rydberg sensor to be competitive with these technologies and have highlighted some of their unique aspects. In particular, being atomic sensors, they hold special appeal as calibration tools since they can be linked directly to fundamental constants and well calculable models. They also only very weakly perturb the measured field which additionally lends to their capabilities as precision sensors. As a relatively new technology, active research is steadily improving their sensitivity and performance with respect to other metrics of interest. While the exact, high-impact application has yet to be conclusively identified for the Rydberg sensor, this work should aid in identifying it.

\begin{acknowledgments}
We thank Fredrik Fatemi and Yuan-Yu Jau for useful discussions. This work was partially supported by the Defense Advanced Research Projects Agency (DARPA).
\end{acknowledgments}

\bibliography{ReceiverComparison}

\end{document}


\renewcommand{\theequation}{S\arabic{equation}}
\renewcommand{\thefigure}{S\arabic{figure}}

\onecolumngrid
\begin{center}
{\Large \textbf{Supplemental Materials for}\\\medskip \textit{Assessment of Rydberg Atoms for Wideband Electric Field Sensing}}
\end{center}
\twocolumngrid

\section{Other Target Rydberg States}
\label{sec:OtherTargets}

As mentioned in Section IV of the main text, the $\state{n}{D}{5/2}$ series of Rydberg states are not the only target states that can be used for the Rydberg sensor. Using the same EIT excitation/readout scheme the $\state{n}{S}{1/2}$ series of states are also accessible via appropriate selection of laser detuning. Due to selection rules, direct optical coupling to the P states requires either single photon excitation \cite{thoumany_optical_2009} or three photon excitation \cite{carr_three-photon_2012}. While the overall strength of the optical coupling to the D-series of Rydberg states is the strongest, leading to the highest SNR signals, these other series provide RF resonances at different frequencies. Incorporating these series into the Rydberg sensor therefore provides a greater range of coverage for the highly sensitive, dipole-allowed transitions than the subset shown in Figure 2b.

\begin{figure}[tb]
    \centering
    \includegraphics[width=\linewidth]{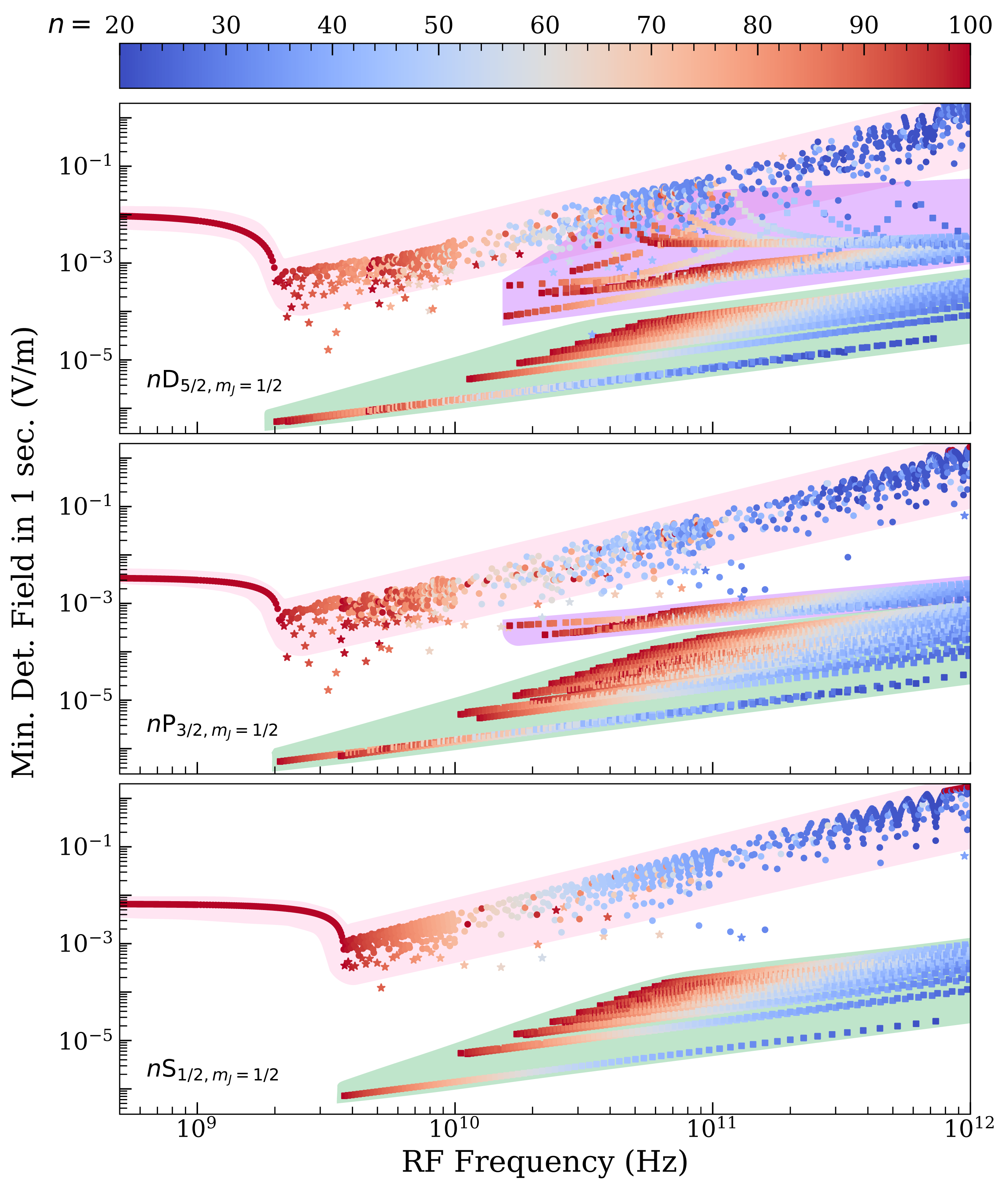}
    \caption{Minimum detectable field in 1 second using other target rubidium Rydberg series. The top panel is the same data for the $\state{n}{D}{5/2}[1/2]$ from Figure 2b. The middle and lower panels show the predicted minimum field for $\state{n}{P}{3/2}[1/2]$ and $\state{n}{S}{1/2}[1/2]$. The green, purple, and pink regions represent different types of couplings as discussed in the text.
    }
    \label{fig:OtherTargets}
\end{figure}

In Figure \ref{fig:OtherTargets} we reproduce the Rydberg sensor minimum detectable field using the $\state{n}{D}{5/2}[1/2]$ series of Rydberg target states along with the corresponding minimum fields for $\state{n}{P}{3/2}[1/2]$ (middle panel) and $\state{n}{S}{1/2}[1/2]$ (lower panel) target states, using the same method described in Section IV A of the main text. The shape and color of each point follows the same convention as that of Figure 2b: the color represents the $n$ that produces the lowest minimum field at that frequency and the shape denotes the SNR scaling of that point with $\Erf$. 

Added to each figure are regional colorings that highlight the different types of RF couplings to the target Rydberg state. Circles in the pink regions represent off-resonant, AC Stark couplings. These points indicate there are no nearby dipole-allowed Rydberg transitions for that particular series of target states. Squares in the green regions indicate the RF frequency is very near or on resonance with a dipole-allowed transition. This type of resonant coupling is very strong and leads to the lowest minimum detectable fields for a particular series. Both of these regions are discussed in detail in Section IV A of the main text. The square points in the purple regions represent dipole-allowed transitions where the coupling is suppressed by $\sim2.5$ orders of magnitude. For the D series target states, these points correspond to couplings with $\state{(n
-\Delta n)}{P}{3/2}$ and $\state{(n+\Delta n)}{F}{7/2}$. For the P series target states, these points correspond to couplings with $\state{(n+\Delta n)}{D}{5/2}$. The mechanism that describes this suppressed coupling is described in the next appendix.

In analyzing these figures, one must first recall that symmetry in the coupling of the Rydberg levels leads to the same transitions on multiple plots. For example, all of the resonant S$\rightarrow$P transitions shown in the bottom panel are present in the middle panel as P$\rightarrow$S transitions with the same magnitude. This property helps in identifying which couplings lead to which lines of resonant minimum field points. Again considering the bottom two panels, the S$\rightarrow$P transitions of the bottom panel are visible in the middle panel as the second series of points in the primary resonance line, slightly offset in minimum field and minimum frequency. The other points of that line are then due to P$\rightarrow$D couplings, which are also present in the top panel.

In general, the minimum detectable field for each target series has similar trends and absolute values, though the exact location of resonances vary from series to series. There are two important difference to note. \textbf{1)} The S$\leftrightarrow$P transitions generally have smaller dipole moments than the other series and are thus less sensitive by a near-unity factor (the minimum field for the $\state{100}{P}{3/2}\rightarrow\state{101}{S}{1/2}$ transition is $\times1.3$ larger than that of $\state{100}{D}{5/2}\rightarrow\state{101}{P}{3/2}$). \textbf{2)} The P and S series have smaller minimum detectable field than the D series in the far-detuned, low frequency regime by factors of 3 and 1.5, respectively. This is because the next nearest transition that would contribute a Stark shift of opposite sign is much further from the lowest $n=100$ resonance for these series. 

\section{Predicting Resonant Rydberg Sensitivities}
\label{sec:radialOverlapTheory}

The strength of the coupling of an RF field to any Rydberg transition is ultimately related to the dipole moment of that transition. 
\begin{equation}
    d=\bra{n\mathrm{L}_{J,m_J}}er\ket{n'\mathrm{L'}_{J',m_J'}}
\end{equation}
This matrix element represents a measure of the amount of overlap of the electron wavefunctions of the two states and can be reduced, using the Wigner-Eckart theorem, to radial and angular terms. The angular terms for all transitions considered in this work are of order 0.5. Significant differences in the dipole moments, observed in the calculations of Section IV of the main text and Section \ref{sec:OtherTargets} of this supplement are due to overlap of the radial wavefunctions. Predicting which transitions will be suppressed (or enhanced) requires some detailed understanding of the radial wavefunction for a Rydberg state.

The radial wavefunction for a Hydrogenic atom can be found by solving the radial portion of the Schr\"odinger equation \cite{gallagher_rydberg_2005}
\begin{equation}\label{eq:RadialSchrodinger}
    \frac{\partial^2\rho(r)}{\partial r^2}+\left[\frac{2}{r}-\frac{1}{{n^*}^2}-\frac{l(l+1)}{r^2}\right]\rho(r)=0
\end{equation}
where $n^*=n-\delta_{nlj}$ is the effective $n$, $\delta_{nlj}$ is the quantum defect, and the wavefunction $\Psi_{nlm}(r,\theta,\phi)=Y_{lm}(\theta,\phi)\rho(r)/r$.

For a Rydberg state with large $n$ the expectation value of the electron radial position is $\braket{r}=(3{n^*}^2-l(l+1))/2$, and gives the portion of the wavefunction with strongest contribution to the wavefunction overlap between two states. The approximate solution of Equation \ref{eq:RadialSchrodinger} near this point is an Airy function of the first kind,
\begin{equation}
    \rho(\braket{r}+\delta r)\propto Ai\left(-{n^*}^{4/3}\left({n^*}^2-\delta r^2-l(l+1)\right)\right)
\end{equation}
which has an oscillatory nature with a large final peak before trending to 0 for $r\rightarrow\infty$. The location of this final peak roughly corresponds to $\braket{r}$ and the dipole moment for these Rydberg transitions strongly depends on the overlap of the final peaks for the wavefunctions of the two states of a transition. To lowest order, the shift of this peak from an $\state{n}{L}{J}$ target state is
\begin{equation}
    \Delta r \approx 3n^{-1/3}(\Delta n-\Delta\delta)
\end{equation}
where $\Delta n$ is the difference in principle quantum numbers and $\Delta\delta$ is the difference in quantum defects between the target state and the coupled state. If $\Delta r$ falls within the full-width half-max (FWHM) of the final peak of the squared Airy function ($\sim1.6$), a large dipole moment for the transition results.

For the S$\leftrightarrow$P, P$\leftrightarrow$D, and D$\leftrightarrow$F transitions considered in this work, $\Delta\delta\approx$\numlist{\pm0.48;\pm1.30;\pm1.33} respectively, with the sign chosen opposite that of $\Delta l$ relative to the target state. Therefore, $\Delta r$ is minimized and the dipole moment increased if the signs of $\Delta n$ and $\Delta l$ for the transition are opposite. This trend is clear in the difference between resonant sensitivities in the green and purple regions of Figure \ref{fig:OtherTargets}, where transitions to states with the same magnitude $\Delta n$, $\Delta l$ can differ by nearly three orders of magnitude in sensitivity depending on relative signs. 

For the S$\leftrightarrow$P transitions this trend is less clear because the absolute value of $\Delta\delta$ is small enough that $\Delta r$ for $\Delta n$, $\Delta l$ of the same sign is closer to the FWHM value. However, these transitions are still an order of magnitude weaker than when $\Delta n$, $\Delta l$ have opposite sign. 

This analysis can also explain why the S$\leftrightarrow$P transitions are generally weaker than the other two series. Limiting ourselves to the most prominent $\Delta n=-1$ transitions, $|\Delta r|\propto0.52$ for S$\rightarrow$P but only $0.3$ for P$\rightarrow$D and D$\rightarrow$F. The increased overlap leads to the larger dipole moments and corresponding lower resonant mimimum detectable fields. 

Finally, this analysis highlights how details of the atomic species used (the quantum defects in this case) can significantly alter the realized sensitivity of the Rydberg sensor. As an example, consider another common species used in Rydberg electrometery: caesium. The values of $\Delta\delta$ for the S$\leftrightarrow$P, P$\leftrightarrow$D, and D$\leftrightarrow$F transitions are approximately \numlist{\pm0.46; \pm1.12; \pm2.44} respectively. Comparing with the rubidium values described above, we can expect a caesium Rydberg sensor to have similar trends due to wavefunction overlap for the S$\leftrightarrow$P transitions, slightly improved overlap for the P$\leftrightarrow$D transitions, and somewhat degraded overlap for the D$\leftrightarrow$F transitions with the strongest resonance corresponding to one of the $|\Delta n|=2$ states.

\section{Experimental Methods}
\label{sec:expMethods}

The experimental configuration is shown in Figure 5 of the main text, and it largely follows the standard Rydberg electrometer configuration found in the literature. 
The probing light is near resonance with the $^{85}$Rb $D2$ transition at \SI{780.24}{\nano\meter} and has a power of \SI{13}{\micro\watt} in a $1/e^2$ beam diameter of \SI{410}{\micro\meter}. Its frequency is controlled via a Direct-Digital Synthesis (DDS) tunable beat-note lock to a master laser that is frequency stabilized via saturated absorption spectroscopy. The \SI{\sim480}{\nano\meter} Rydberg coupling light, with a $1/e^2$ beam diameter of \SI{380}{\micro\meter} and typical power of \SI{\sim500}{\milli\watt}, is frequency stabilized to an ultra-low expansion (ULE) reference cavity. The RF fields are applied using a waveguide horn antenna, with polarization parallel to that of the light. The atomic response is measured by sweeping the probe light frequency through resonance sufficiently slowly (typically \SI{67}{\mega\hertz\per\second}) to ensure a steady-state probing regime. Experimental timing and control is performed using the \texttt{labscript} suite \cite{starkey_scripted_2013}.

The transmitted probe light is measured using an optical homodyne method similar to that used in Refs.~\cite{mohapatra_giant_2008,kumar_atom-based_2017} which allows for precise, photon-shot-noise limited measurements in both the phase and amplitude quadratures. Our implementation follows that of Ref.~\cite{cox_generating_2015} where we simultaneously measure an optical heterodyne with a sideband of the probe to stabilize the relative beam path phase. This method has the advantage of an easy change between measurement quadratures without the need for a second reference laser. In this work, all data is taken in the amplitude quadrature.

In our implementation the \SI{780}{\nano\meter} laser light is initially separated into probe and local oscillator (LO) paths, and frequency shifted up and down \SI{78.6}{\mega\hertz}, respectively, by separate acousto-optic modulators (AOMs). These AOMs are also used to stabilize the power in the optical beams. An electro-optic phase modulator (EOM) then imparts \SI{157.2}{\mega\hertz} sidebands on the probe beam, such that the lower sideband is at the same frequency as the LO beam, facilitating homodyne detection, see Fig.~5c of the main text. The relative beam path phase is stabilized using the simultaneous \SI{157.2}{\mega\hertz} heterodyne signal between LO and probe carrier frequency. Thus, the balanced photodetector output has two signals of interest, a DC-coupled homodyne signal carrying the spectroscopic information, and a heterodyne beat-note at \SI{157.2}{\mega\hertz}, which provides the correction signal that is fed back to the EOM for path stabilization. Note that the drive power of the EOM is relatively low, meaning most of the probe power is in the carrier. This minimizes the influence of the upper sideband and affords precise control of the probe power in the lower sideband while maintaining higher carrier optical power for the heterodyne signal.

This system requires some care in the details of its configuration. First, the probe carrier and upper sideband should be blue detuned from atomic resonance to avoid their interaction with the EIT signal of interest. Second, the path lengths should be passively balanced and the RF signals to the AOMs, EOM, and heterodyne mixer's LO should be phase coherent in order to reduce the impact of the RF source's phase noise. We obtained phase coherence by deriving all RF signals from the same DDS synthesizer (that was externally clocked to a low-noise \SI{100}{\mega\hertz} reference oscillator, which was in turn stabilized to a rubidium reference), and matching the cable delays for the AOM and EOM drives. Due to the slow acoustic velocity in the AOMs (\SI{4}{\milli\meter\per\micro\second}), particular attention must be taken to ensure the probe and reference beams are the same distance from the transducer. The optical path length between the LO and probe should also be balanced to limit the influence of phase noise from the probe laser itself.

\bibliography{ReceiverComparison}